\documentclass[useAMS,usenatbib]{mn2e}
\usepackage[fleqn]{amsmath}
\usepackage{amssymb}
\usepackage{txfonts}
\usepackage{graphicx}
\usepackage{multirow}
\usepackage{dcolumn}
\usepackage{array}
\usepackage{xcolor}
\usepackage{float}
\title[]{What mechanisms dominate the activity of Geminid Parent (3200) Phaethon?}

\author[LiangLiang Yu et al.]{LiangLiang Yu$^{1}$\thanks{yullmoon@live.com}, Wing-Huen Ip$^{1,2}$, Tilman Spohn$^{3}$\\
$^{1}$Space Science Institute, Macau University of Science and Technology, Taipa, Macau; \\
$^{2}$Institute of Astronomy, National Central University, Jhongli, Taoyuan City 32001, Taiwan; \\
$^{3}$Institute of Planetary Research, German Aerospace Center (DLR), Rutherfordstra$\beta$e 2, 12489 Berlin}

\begin{document}
\date{Accepted 2018 October 28. Received 2018 October 5; in original form 2018 July 18}

\pagerange{\pageref{firstpage}--\pageref{lastpage}} \pubyear{2018}

\maketitle

\label{firstpage}

\begin{abstract}
A long-term sublimation model to explain how Phaethon could provide the Geminid stream
is proposed. We find that it would take $\sim6$ Myr or more for Phaethon to lose all of
its internal ice (if ever there was) in its present orbit. Thus, if the asteroid moved
from the region of a 5:2 or 8:3 mean motion resonance with Jupiter to its present orbit
less than $1$ Myr ago, it may have retained much of its primordial ice. The dust mantle
on the sublimating body should have a thickness of at least $15$ m but the mantle could
have been less than $1$ m thick $1000$ years ago. We find that the total gas production
rate could have been as large as $10^{27}\rm~s^{-1}$ then, and the gas flow could have
been capable of lifting dust particles of up to a few centimeters in size. Therefore,
gas production during the past millennium could have been sufficient to blow away enough
dust particles to explain the entire Geminid stream. For present-day Phaethon, the gas
production is comparatively weak. But strong transient gas release with a rate of
$\sim4.5\times10^{19}\rm~m^{-2}s^{-1}$ is expected for its south polar region when
Phaethon moves from $0^\circ$ to $2^\circ$ mean anomaly near perihelion. Consequently,
dust particles with radii of $<\sim260~\mu m$ can be blown away to form a dust tail.
In addition, we find that the large surface temperature variation of $>600$ K near
perihelion can generate sufficiently large thermal stress to cause fracture of rocks
or boulders and provide an efficient mechanism to produce dust particles on the surface.
The time scale for this process should be several times longer than the seasonal thermal
cycle, thereby dominating the cycle of appearance of the dust tail.
\end{abstract}


\section{Introduction}
(3200) Phaethon, discovered by S.F. Green in October 1983 from data of the Infrared
Astronomical Satellite (IRAS) \citep{Green1983}, is an intriguing Apollo-group
Near-Earth Object with an unusual elliptical ($e\sim0.89$) orbit that crosses
the orbits of Mars, Earth, Venus and Mercury. The dynamical characteristics of
Phaethon and of the Geminid meteoroids suggest that Phaethon may be the source of
the Geminid meteor stream, with some of the meteoroids being ejected from Phaethon
at its perihelion during the past $\sim1000$ years \citep{Gustafson1989}. If true,
Phaethon would have had a comet-like activity, at least for the past millennium.

Most meteoroid streams are of cometary origin \citep{Jenniskens2008} where mass
losses are driven by the sublimation of near-surface volatiles. Phaethon, judging
from its dynamical characteristics, is an asteroid and is not likely to retain
near-surface volatiles for a long time because of its periodically high surface
temperature at perihelion. Its present perihelion distance of $\sim0.14$ AU
suggests surface temperatures as high as $\sim1000$ K, thus raising questions
about the mechanisms for continued significant mass loss from Phaethon.

\citet{Jewitt2013} detected Phaeton's tail in data taken by the STEREO spacecraft
during two perihelion passages in 2009 and 2012, respectively, thereby confirming
Phaeton's previously predicted activity. \citet{Hui2017} again observed the dust tail
during Phaethon's perihelion passage in 2016. The recurring of the tail at perihelion
implies that it should have a periodic cause associated with the asteroid's orbital
position. \citet{Jewitt2013} and \citet{Jewitt2015} explained the tails as being due to
a combination of radiation pressure sweeping particles away and thermal disintegration
of the asteroid surface caused by rotational thermal stress or thermal desiccation
cracking, which seems to be the most convincing mechanisms at present. However, there
still remains questions in this scenario.

First, it should be expected that similar small bodies with similar small perihelion
distances as Phaethon should also have such dust tails, and may be expected to be
associated with meteor streams as well. But \citet{Jewitt2013b} show that other
asteroids with similar small-perihelion distance do not have tails as Phaethon does.

Second, all of the three observations \citep{Jewitt2013,Hui2017} reported that the tail
suddenly appeared at Phaeton's perihelion mean anomaly $=0^\circ\sim1^\circ$ and continued
for a very short time of $\sim2$ days \citep{Li2013}. This appears more like a short-term
outburst rather than the result of continuous solar radiation sweeping.

Moreover, both \citet{Jewitt2013} and \citet{Hui2017} argued that the particles in Phaethon's
tail were mainly small particles with sizes of $\sim1~\mu\rm{m}$. These would be quickly
swept away by the solar radiation pressure and thus can hardly account for the Geminids.
The Geminid meteor stream consist of much larger particles with sizes ranging from
$\sim10~\mu\rm{m}$ to $\sim1$ cm \citep{Yanagisawa2008,Borovicka2010,Blaauw2017}. These
unresolved questions imply that other mechanisms than those cited above may be responsible
for the observed mass loss from Phaethon.

Therefore, it has been speculated that Phaethon may still contain relatively pristine
volatiles buried below the surface and that gaseous outburst near perihelion could eject
large dust particles to supply the Geminid meteor stream \citep{Boice2015,Boice2017}.
This idea is supported by the results of \citet{Licandro2007} that Phaethon's spectrum is
consistent with that of B-type asteroids, similar to that of CI/CM meteorites altered by
aqueous activity and of hydrated minerals. \citet{deLeon2010} suggested that Phaethon may
be a fragment of (2) Pallas because of their compositional similarities and dynamical
connections. The hypothesis of a dynamical connection between Phaethon and Pallas is
further supported by \citet{Todorovi2018}, who finds a significant probability of $>40\%$
for a fast delivery (on a time scale of $\sim1$ Myr) of Phaethon from the region of a 5:2
or 8:3 mean motion resonance with Jupiter to its present orbit.

Generally, B-type small bodies, a sub-class of the wider C-type, are mostly believed to
be primitive, volatile-rich remnants from the early Solar System \citep{Rivkin2002}. The
discovery of main-belt comets by \citet{Hsieh2006} further supports the idea that main-belt
objects could still have retained ice. Therefore,  Pallas could be an icy body, and if
Phaethon were an icy fragment of Pallas, it could have at least started in its present
orbit with ice buried underneath its surface.

In this work, we aim to investigate whether ice could have been retained below the
surface of Phaethon assuming that the latter is a fragment of Pallas, and whether
sublimation-driven activity could explain the Phaethon-Geminids connection, as well
as the tail observed during recent perihelion passages.

\section{Thermal Model of Icy Bodies}
\subsection{Dust-ice two-layer system}
\begin{figure}
\includegraphics[scale=0.58]{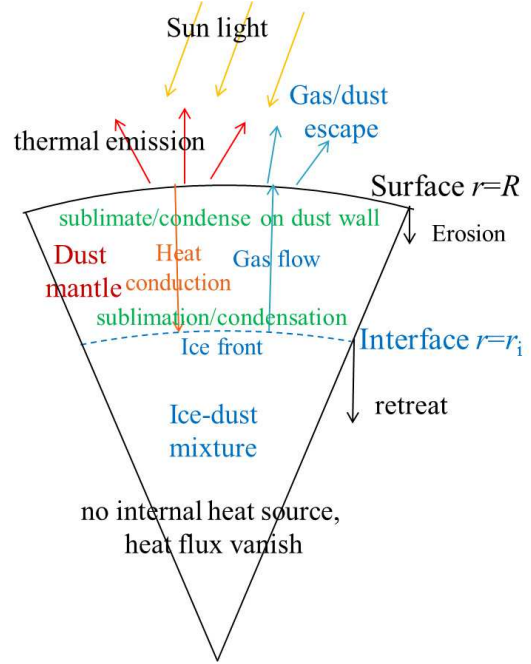}
  \centering
  \caption{Dust layer on top of the ice-dust interior of an icy body.
  }\label{twolayer}
\end{figure}

Consider a B-type asteroid of radius $R$ with an initial ice/dust mass ratio of $\chi_0$.
If such a body is transferred to an orbit close enough to the sun where the surface
temperature gets high enough for the sublimation of ice to occur, then a dust mantle
with porosity $\phi$ and tortuosity $\varsigma$ will gradually grow on its surface.
And the ice front, the interface between the dust mantle and the icy interior at depth
$h_{\rm i}$ (or raidus $r_{\rm i}$), will gradually retreat towards the interior
(compare Figure \ref{twolayer}).

In this system, there are three thermo-physical processes --- heat flow,
Knudsen gas flow and sublimation/condensation, which can be described by
the equations of thermal diffusion
\begin{equation}
\rho c\frac{\partial T}{\partial t}+\nabla\cdot\vec{q}
+c_{\rm g}\hat{m}\vec{j}\cdot\nabla T=Q~,
\label{heatd}
\end{equation}
gas diffusion
\begin{equation}
\frac{\partial n}{\partial t}+\nabla\cdot\vec{j}=J\xi~,
\end{equation}
and sublimation/condensation
\begin{equation}
\frac{{\rm d}\varphi}{{\rm d}t}=-\frac{\hat{m}J\xi}{\rho_{\rm i}}~.
\label{subl}
\end{equation}
In Equation (\ref{heatd}) - (\ref{subl}), $\rho$ is the solid mass density, $c$ is solid
specific heat capacity, $T$ is temperature, a function of radial distance $r$ from the
center and latitude, $c_{\rm g}$ is the gas specific heat capacity, $\hat{m}$ is the gas
molecular mass, $n$ is the number density of gas molecules, $\varphi$ is the volume fraction
of ice, $\xi$ is the surface to volume ratio of the dust grain matrix,
\begin{equation}
\vec{q}=-\kappa\nabla T
\end{equation}
is the heat flux,
\begin{equation}
\vec{j}=-\beta\nabla n
\end{equation}
is gas diffusion flux,
\begin{equation}
Q=-\Delta H\hat{m}J\xi
\end{equation}
is the latent heat exchange caused by sublimation/condensation, where $\Delta H$
is the enthalpy of ice sublimation and is a function of temperature,
\begin{equation}
J=\frac{1}{4}\tilde{v}_{\rm th}(n_{\rm E}-n)
\end{equation}
is the rate of sublimation/condensation,
\begin{equation}
\tilde{v}_{\rm th}=\sqrt{\frac{8k_{\rm B}T}{\pi \hat{m}}}
\end{equation}
is the mean thermal velocity, $n_{\rm E}$ is the equilibrium (or saturation)
number density, which is calculated from the temperature via the ideal gas law
and the integral form of Clausius-Clapeyron equation
\begin{equation}
n_{\rm E}=\frac{P_{\rm E}}{k_{\rm B}T}~,~
P_{\rm E}=P_{0}\exp\left(-\frac{\Delta H_{\rm m}}{R_{\rm g}}
\left(\frac{1}{T}-\frac{1}{T_0}\right)\right)
\label{nE}
\end{equation}
where $P_{0}=611\rm~Pa$, $T_0= 273.15\rm~K$, $\Delta H_{\rm m}=\Delta H\hat{m}N_{\rm A}$
is the mol enthalpy of ice sublimation, $R_{\rm g}$ is the gas constant, and $N_{\rm A}$
is Avogadro's number.
In the case of an approximately constant sublimation temperature of about $210\rm~K$,
$\Delta H$ can be taken as constant:
\[\Delta H\approx2.83*10^6\rm~Jkg^{-1}.\]

The gas flow in the dust mantle can be treated as  "Knudsen diffusion" (diffusion
of a rarified gas), where the Knudsen diffusion coefficient $\beta$, for porous media
can be estimated from \citep{Schorghofer2008}
\begin{equation}
\beta=\left(\frac{\pi}{8+\pi}\frac{\phi}{1-\phi}\frac{b}{\varsigma}\right)\tilde{v}_{\rm th},
\label{beta}
\end{equation}
where $b$ is the mean grain radius, $\phi$ is porosity and $\varsigma$ is tortuosity.

The thermal conductivity $\kappa$ of the dust mantle as a function of temperature $T$,
mean grain radius $b$ and porosity $\phi$ can be modeled following \citet{Gundlach2013}:
\begin{eqnarray}
\kappa(T,b,\phi)&=&\kappa_{\rm solid}
\left(\frac{9\pi}{4}\frac{1-\mu^{2}}{E}\frac{\gamma(T)}{b}\right)^{1/3}
\cdot f_{1}e^{f_{2}(1-\phi)}\cdot\chi
\nonumber\\
&&+8\sigma\epsilon T^{3}\frac{e_{1}\phi}{1-\phi}b ~,
\label{ktrphi}
\end{eqnarray}
where $\kappa_{\rm solid}$ is the thermal conductivity of the solid material, $\mu$ is
Poisson's ratio, $E$ is Young's modulus, $\gamma(T)$ is the specific surface energy,
$\epsilon$ is the emissivity of the material, and $f_1$, $f_2$, $\chi$ and $e_1$ are
best-fit coefficients. For more details, we refer the reader to \citet{Gundlach2013}.

\subsection{Long-term sublimation of buried ice}
We are considering the long-term evolution of Phaethon, that is the evolution on a
time scale much longer than the orbital period. In this case, the thickness of the
dust mantle $h_i$ by far exceeds the seasonal thermal skin depth $l_{\rm sst}$
\[h_i=R-r_i\gg l_{\rm sst}\sim 1\rm~m,\]
with $r_i$ the radial distance to the bottom of the dust mantle, and the model can
be significantly simplified by taking the following assumptions:

(1) Although the surface temperature varies periodically due to rotation and orbital
movement, the subsurface temperature below several seasonal thermal skin depths
will be in a quasi-static equilibrium with the average solar insolation and thermal
emission from the surface.

(2) The heat flow in the dust mantle from the surface to the ice sublimation front
is mainly consumed by the latent heat of sublimation there.

(3) The difference between the subsurface temperature $\tilde{T}_0$ in equilibrium with
the insolation and the temperature $T_{\rm i}$ of the ice front can be considered constant.
This temperature difference continuously drives sublimation at the ice front.

The equilibrium subsurface temperature $\tilde{T}_0$ can be estimated by
assuming radiation equilibrium at the surface
\begin{equation}
\tilde{T}_0=\left[\frac{(1-A_{\rm B})\tilde{F}}{\varepsilon\sigma}\right]^{1/4},
\end{equation}
where $A_{\rm B}$ is the Bond albedo, $\varepsilon$ is the average thermal emissivity,
and $\tilde{F}$ is the annual average solar insolation. $A_{\rm B}$ can be estimated from
the geometric albedo $p_{\rm v}$ via
\begin{equation}
A_{\rm B}=(0.29+0.684G)p_{\rm v},
\end{equation}
where $G$ is the slope parameter in the H,G magnitude system of \citet{Bowell}.
Because of rotation and the incidence angle of insolation varying with latitude $\theta$,
both the annual average solar insolation $\tilde{F}$ and $\tilde{T}_0$ will vary with latitude.
Therefore, it is necessary to determine $\tilde{F}=\tilde{F}(\theta)$ for the actual
orbit and rotation state.

To calculate the temperature $T_{\rm i}$ at the ice front ($r=r_i$), we start with
the energy balance equation at the ice front (in thermal equilibrium)
\begin{equation}
\nabla\cdot\vec{q}=Q=-\Delta H\hat{m}J\xi,
\end{equation}
from which we obtain
\begin{equation}
q_{\rm i}(r_{\rm i})=-\Delta H\hat{m}J,
\label{qi}
\end{equation}
which implies that the heat flow at the ice front is consumed by ice sublimation.
For the heat flow at radius $r$ in the dust mantle, energy conservation further gives
\begin{equation}
q=-\kappa\frac{{\rm d}T}{{\rm d}r}\approx\frac{q_{\rm i}r_{\rm i}^2}{r^2}.
\label{qqi}
\end{equation}

We assume that the sublimation flux $J$ diffuses from the ice front to the surface.
From mass conservation, we find the gas flow flux at a radial distance $r$ to be
\begin{equation}
j=-\beta\frac{{\rm d}n}{{\rm d}r}\approx\frac{Jr_{\rm i}^2}{r^2}.
\label{jJ}
\end{equation}

Integrating both sides of Equation (\ref{qqi}) and Equation (\ref{jJ}), we obtain
\[T\Big|^{r=r_{\rm i}}_{r=R}\approx\int^{r_{\rm i}}_{R}\frac{q_{\rm i}}{\kappa}
{\rm d}\left(\frac{1}{r}\right),\]
\[n\Big|^{r=r_{\rm i}}_{r=R}\approx\int^{r_{\rm i}}_{R}\frac{J}{\beta}
{\rm d}\left(\frac{1}{r}\right),\]
which further gives,
\begin{equation}
T_{\rm i}-\tilde{T}_0\approx r_{\rm i}\left(1-\frac{r_{\rm i}}{R}\right)\frac{q_{\rm i}}{\tilde{\kappa}},
\label{TTi}
\end{equation}
\begin{equation}
n_{\rm i}-n_0\approx r_{\rm i}\left(1-\frac{r_{\rm i}}{R}\right)\frac{J}{\tilde{\beta}},
\label{nni}
\end{equation}
assuming $\tilde{\kappa}$ and $\tilde{\beta}$ to be constant in the dust mantle with
their average values calculated from Equation (\ref{beta}) and (\ref{ktrphi}).

For small bodies, the gravity is too low to prevent thermal escape of the water vapor
molecules at the surface, thus we take
\[n_0\approx0.\]
And at the ice front, because of sublimation,
\[J=\frac{1}{4}\tilde{v}_{\rm th}(T_{\rm i})(n_{\rm E}(T_{\rm i})-n_{\rm i}),\]
from which we obtain using Equation (\ref{TTi}) and (\ref{nni})
\begin{equation}
n_{\rm i}=\frac{r_{\rm i}\left(1-\frac{r_{\rm i}}{R}\right)}{
r_{\rm i}\left(1-\frac{r_{\rm i}}{R}\right)+4\beta/\tilde{v}_{\rm th}(T_{\rm i})}
n_{\rm E}(T_{\rm i}),
\end{equation}
\begin{equation}
J=\frac{\beta n_{\rm E}(T_{\rm i})}{
r_{\rm i}\left(1-\frac{r_{\rm i}}{R}\right)+4\beta/\tilde{v}_{\rm th}(T_{\rm i})}.
\label{Jhi}
\end{equation}
Moreover, combining Equation (\ref{qi}), Equation (\ref{TTi}) and Equation (\ref{nni}), we find
\begin{equation}
\tilde{T}_0=T_{\rm i}+\frac{\Delta H\hat{m}\tilde{\beta}n_{\rm i}(T_{\rm i})}{\tilde{\kappa}},
\end{equation}
which enables us to calculate the temperature $T_{\rm i}$ of the ice front from the
equilibrium subsurface temperature $\tilde{T}_0$.

Here, we should note that when the ice front radius
\[r_{\rm i}\rightarrow0~{\rm or}~r_{\rm i}\rightarrow R,\] then
\[r_{\rm i}\left(1-\frac{r_{\rm i}}{R}\right)\rightarrow0,\]
implying that the number density $n_{\rm i}$, the temperature $T_{\rm i}$,
and the sublimation rate $J$ at the ice front tend towards
\[n_{\rm i}\rightarrow0,T_{\rm i}\rightarrow\tilde{T}_0,{\rm and},
J\rightarrow\frac{1}{4}\tilde{v}_{\rm th}(\tilde{T}_0)n_{\rm E}(\tilde{T}_0).\]

However, the Knudsen diffusion coefficient $\beta$ of a rarified gas in the
dust mantle generally satisfies
\[\frac{4\tilde{\beta}}{\tilde{v}_{\rm th}(T_{\rm i})}\sim b \ll 1~{\rm m}.\]
Thus, when the ice front radius
\[1~{\rm m}<r_{\rm i}<R-1~{\rm m}\]
the condition
\[\frac{4\tilde{\beta}}{\tilde{v}_{\rm th}(T_{\rm i})}
\ll 1<r_{\rm i}\left(1-\frac{r_{\rm i}}{R}\right)\]
will always be true, so that in most of the interior of the small body,
\begin{equation}
n_{\rm i}\approx n_{\rm E}(T_{\rm i}),
\end{equation}
\begin{equation}
\tilde{T}_0=T_{\rm i}+\frac{\Delta H\hat{m}\tilde{\beta}n_{\rm E}(T_{\rm i})}{\tilde{\kappa}},
\label{Ti}
\end{equation}
and
\begin{equation}
J\approx\frac{\tilde{\beta}n_{\rm E}(T_{\rm i})}{r_{\rm i}\left(1-\frac{r_{\rm i}}{R}\right)}.
\label{Jhii}
\end{equation}

According to Equation (\ref{Jhi}) and Equation (\ref{Jhii}), when the ice front
radius reaches the value $r_{\rm i}=R/2$, the term
\[r_{\rm i}\left(1-\frac{r_{\rm i}}{R}\right)=\frac{R}{4},\]
will attend a maximum value while the sublimation rate $J$ attends a minimum value
\begin{equation}
J_{\rm min}=\frac{4\tilde{\beta}n_{\rm E}(T_{\rm i})}{R}.
\label{mJmin}
\end{equation}
Therefore, as the ice front $r_{\rm i}$ retreats from the surface to $R/2$,
the sublimation rate $J$ decreases from
\[\frac{1}{4}\tilde{v}_{\rm th}(\tilde{T}_0)n_{\rm E}(\tilde{T}_0)
\rightarrow\frac{4\tilde{\beta}n_{\rm E}(T_{\rm i})}{R}.\]
For even deeper sublimation, when $r_{\rm i}$ moves from $R/2$ to the body center,
the sublimation rate $J$ increases from
\[\frac{4\tilde{\beta}n_{\rm E}(T_{\rm i})}{R}\rightarrow
\frac{1}{4}\tilde{v}_{\rm th}(\tilde{T}_0)n_{\rm E}(\tilde{T}_0),\]
until all ice is exhausted.

With the sublimation rate $J$, the rate of retreat of the buried ice front
can be calculated from
\begin{eqnarray}
\frac{{\rm d}r_{\rm i}}{{\rm d}t}&=&-\frac{\hat{m}J}{(1-\phi)\rho_{\rm d}\chi_0}
\approx-\frac{1}{r_{\rm i}\left(1-\frac{r_{\rm i}}{R}\right)}
\frac{\hat{m}\tilde{\beta}n_{\rm E}}{(1-\phi)\rho_{\rm d}\chi_0}
\nonumber\\
&\equiv&-\frac{R_r}{r_{\rm i}\left(1-\frac{r_{\rm i}}{R}\right)},
\label{drdt}
\end{eqnarray}
or
\begin{equation}
\frac{{\rm d}h_{\rm i}}{{\rm d}t}=-\frac{{\rm d}r_{\rm i}}{{\rm d}t}
\approx\frac{R_r}{r_{\rm i}\left(1-\frac{r_{\rm i}}{R}\right)}
=\frac{R_r}{h_{\rm i}\left(1-\frac{h_{\rm i}}{R}\right)},
\label{dhdt}
\end{equation}
where $R_r$ is defined as
\begin{eqnarray}
R_r&\equiv&\frac{\hat{m}\tilde{\beta} n_{\rm E}}{(1-\phi)\rho_{\rm d}\chi_0}
\nonumber\\
&=&\left(\frac{\pi}{8+\pi}\frac{\phi}{(1-\phi)^2}\frac{b}{\rho_{\rm d}\chi_0\varsigma}\right)
\hat{m}\tilde{v}_{\rm th}n_{\rm E}(T_{\rm i}),
\end{eqnarray}
in which
\[\tilde{v}_{\rm th}=\frac{1}{2}(\tilde{v}_{\rm th}(\tilde{T}_0)+\tilde{v}_{\rm th}(T_{\rm i})).\]
$R_r$ can be understood as the characteristic rate at which the sublimation front retreats,
we thus name it "retreating rate". Because the saturation number density $n_{\rm E}$ depends
exponentially on temperature (compare Equation (\ref{nE})), the value of $R_r$ is mainly a
function of the ice front temperature $T_{\rm i}$.

If the erosion of the surface due to gas drag can be ignored on the time scale considered,
$R=\rm const.$
In the case of a stable orbit and rotation state, $\tilde{T}_0$ and $T_{\rm i}$ will be
constant in time, so that, in addition,
$R_r=\rm const.$
By integrating both sides of Equation (\ref{drdt}) or Equation (\ref{dhdt}), we obtain
\begin{equation}
\left(\frac{r_{\rm i}^2}{2}-\frac{r_{\rm i}^3}{3R}\right)\approx
\left(\frac{r_0^2}{2}-\frac{r_0^3}{3R}\right)-R_rt,
\label{rit}
\end{equation}
or
\begin{equation}
\left(\frac{h_{\rm i}^2}{2}-\frac{h_{\rm i}^3}{3R}\right)\approx
\left(\frac{h_0^2}{2}-\frac{h_0^3}{3R}\right)+R_rt,
\label{hit}
\end{equation}
where $r_0$ and $h_0$ are the initial radius and depth to the ice front, respectively, and
$r_{\rm i}$ and $h_{\rm i}$ are the radius and depth to the ice front, respectively, as
functions of time $t$.

As a special case, if the depth to the ice front $h_{\rm i}\ll R$, we have
\begin{equation}
\frac{{\rm d}h_{\rm i}}{{\rm d}t}
\approx\frac{R_r}{h_{\rm i}\left(1-\frac{h_{\rm i}}{R}\right)}
\approx\frac{R_r}{h_{\rm i}},
\end{equation}
from which
\begin{equation}
h_{\rm i}\approx\sqrt{h_0^2+2R_rt}.
\end{equation}

Moreover, if the initial radial distance from the center of the ice front is $r_0=R$,
then the time scale to lose all the ice ($r_{\rm i}=0$) is
\begin{equation}
t_{\rm L}=\frac{R^2}{6R_r}.
\end{equation}

\section{Long-term thermal evolution}
\subsection{Could Phaethon once have contained ice?}
To discuss whether or not Phaethon may have retained ice through its history to the
present day, we will have to know how and where the asteroid originated. If Phaethon
formed as a fragment of Pallas, blown off by a relatively recent impact, for example,
then the question would be whether Pallas could still have contained ice at the time
of the recent Phaeton-forming collision after billions of years of evolution.

Pallas is a B-type asteroid most likely formed in the solar nebula beyond the
snow line \citep{Sasselov2000} of the early solar system. With an effective
diameter of $\sim545\rm~km$, Pallas is large enough so that its orbit and
rotation can be stable even on the time scale of the age of the solar system.
Thus, Equation (\ref{hit}) can be applied to estimate the depth to any remaining
ice in the interior. The orbital parameter values for Pallas can be found in the
JPL Small-Body database and the rotation parameters have been well constrained by
\citet{Carry2010} with a rotation period of $7.8134\rm~hr$ and angles to the rotation
axis of ($-16.0^{\circ}$, $30.0^\circ$) in the Ecliptic J2000 reference frame,
implying a high axial tilt of $84\pm5^\circ$. Reasonable physical parameter
values for the dust mantle are listed in Table \ref{papa}.

\begin{table}
 \centering
 \renewcommand\arraystretch{1.3}
 \caption{Assumed physical parameters of the dust mantle of Pallas.}
 \label{papa}
 \begin{tabular}{@{}lcc@{}}
 \hline
 Property & Value \\
 \hline
 Grain density $\rho_{\rm d}$  & $3000\rm~kgm^{-3}$    \\
 Mean Grain radius $b$         &  $100\rm~\mu m$        \\
 Tortuosity $\varsigma$   &  2         \\
 Porosity $\phi$          &  0.5       \\
 Ice/dust ratio  $\chi_0$ &  0.1$\sim$0.2     \\
% Heat conductivity      & $\sim5\times10^{-4}$    \\
 \hline
\end{tabular}
\end{table}

\begin{figure}
\includegraphics[scale=0.58]{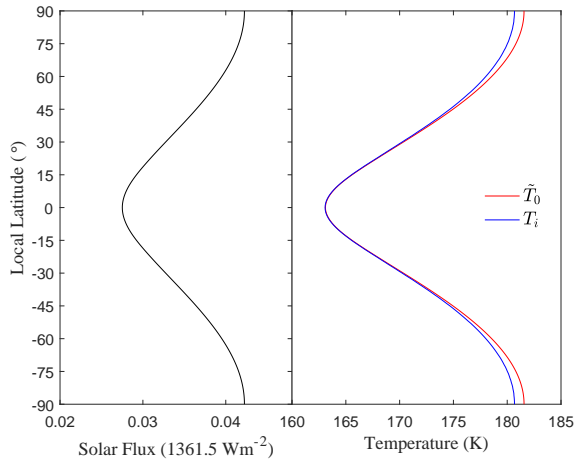}
  \centering
  \caption{Left panel: Annual mean solar insolation as a function of  local latitude on Pallas;
  Right panel: Equilibrium subsurface temperature $\tilde{T_0}$ and ice sublimation front temperature
  $T_{\rm i}$ as functions of local latitude, calculated assuming a geometric albedo of $p_{\rm v}=0.1$.
  }\label{PallasAFAT}
\end{figure}

The rotation axis of Pallas nearly lies in the orbital plane causing the polar regions to receive
more solar insolation than the equatorial regions (left panel in Figure \ref{PallasAFAT} ). The
right panel in Figure \ref{PallasAFAT} shows the equilibrium subsurface temperature $\tilde{T_0}$
on each local latitude of Pallas and the corresponding ice front temperature $T_{\rm i}$ derived
from Equation (\ref{Ti}). The equatorial subsurface temperature is smaller than the subsurface
temperature at higher latitudes, implying that the loss rate of ice on the equator should be
smaller than the loss rate at high latitudes.

\begin{figure}
\includegraphics[scale=0.58]{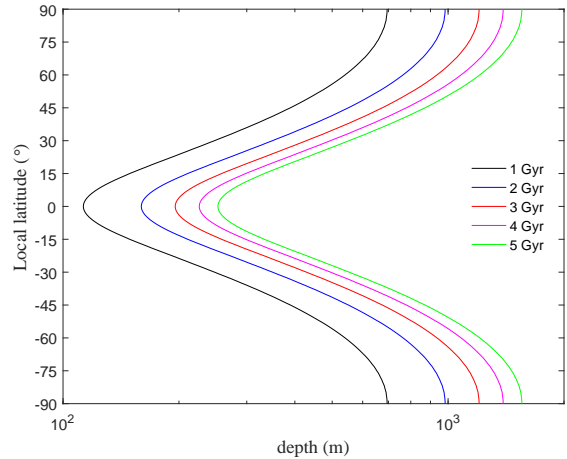}
  \centering
  \caption{Depth to the ice front in Pallas calculated from the present model}
  \label{PallasHt}
\end{figure}

Given the ice sublimation front temperature, the inward retreating rate $R_r(\theta)$
can be estimated for each latitude, and then the secular evolution of the ice front can
be evaluated as shown in Figure \ref{PallasHt}. Along the equator of Pallas, the inward
retreating rate $R_r$ of the ice front is at a minimum value of $\sim10^{-13}\rm~m^2s^{-1}$,
suggesting that the time for losing all ice would be
\[t_{\rm L}\sim\frac{R^2}{6R_r}\sim10^7\rm~Gyr.\]
This time is much longer than the age of the solar system. Moreover, from Figure \ref{PallasHt},
we find that even on Gyr time scales, the depth to the ice front would still be within
$\sim100-1000$ meters of the surface, suggesting that the dust mantle would have a thickness
in the range of several hundred meters at most (because we do not consider any losses of dust
mantle due to gas escape). This thickness of the dust mantle is much smaller than the size
of Phaethon. Thus, if Phaethon were a fragment of Pallas formed by a giant impact, it would
be likely that the bulk of Phaethon would contain ice, suggesting that Phaethon could
have started as an icy body.

\subsection{Can Phaethon have retained ice to the present day?}
To answer the question of whether or not Phaethon can still retain ice in its interior
(assuming it formed as an icy body), we would have to know its orbital evolution. We
would then be able to model its thermal evolution as a consequence of its time varying
insolation and trace the secular inward motion of the possible ice front. Unfortunately,
it is difficult to exactly reconstruct the orbital evolution of a small body like Phaethon
because the orbit was likely not stable. Moreover, if we assume Phaethon to be a fragment
of Pallas, its orbit may have changed so much that its thermal history becomes almost
impossible to reconstruct. However, because Phaethon is in a retrograde orbit, it is
conceivable that its semi-major axis is continuously decreasing because of the Yarkovsky
effect. The average subsurface temperature should then be increasing with time. Thus the
present equilibrium subsurface temperature, which can be easily calculated from the present
orbit and rotation, can be used to place an upper limit on the past subsurface temperature.

We use the orbital parameters of Phaethon as published in the JPL Small-Body database
and the rotation parameters derived by \citet{Hanus2016} --- rotation period
$P_{\rm rt}=3.604\rm~hr$ and orientation ($\lambda=319^\circ$, $\beta=-39^\circ$)
in the Ecliptic coordinate system.

The physical parameter values for the dust mantle on Phaethon should be slightly different
from those for Pallas because Phaethon has a much larger average surface thermal inertia of
$\sim600$ $\rm Jm^{-2}s^{-0.5}K^{-1}$ \citep{Hanus2016}, suggesting a larger density and/or
larger mean grain size. The physical properties of the Geminids may provide useful information
on Phaethon's dust mantle if we assume that the Geminids do come from Phaethon.
\citet{Borovicka2010} found the bulk density of the Geminids to be $\sim2600~{\rm kgm^{-3}}$,
indicating a "micro-porosity" of $\sim13\%$ for a material density of $\sim3000~{\rm kgm^{-3}}$
(or anything reasonably close to that). Here we use $\sim3000~{\rm kgm^{-3}}$ for the material
density of dust grains that form the dust mantle, but assume a "macro-porosity" of $\sim40\%$,
somewhat smaller than that used for Pallas, considering that sintering may have reduced the
dust mantle porosity.

The recent polarization measurements of \citet{Ito2018} suggest that grains of $\sim$150 $\mu$m
radius may characterize Phaethon's surface. Considering that the grains on the very surface may
be easier to fracture due to space weathering or thermal stress, whereas grains in the subsurface
dust mantle may aggregate and grow by sintering, we expect the average grain size on the surface
to be smaller than the mean grain size in the subsurface dust mantle, if the formed dust mantle
is enough thick ($>1$ m for example). Therefore, we assume a mean  grain size of $\sim500$ $\mu$m
for the dust mantle. We list our chosen parameter values in Table (\ref{phpa}).

\begin{table}
 \centering
 \renewcommand\arraystretch{1.3}
 \caption{Assumed physical parameters of Phaethon's dust mantle.}
 \label{phpa}
 \begin{tabular}{@{}lcc@{}}
 \hline
 Property & Value \\
 \hline
 Grain density $\rho_{\rm d}$  & $3000\rm~kgm^{-3}$  &  \\
 Mean Grain radius $b$        &  $500\rm~\mu m$      &  \\
 Tortuosity $\varsigma$  &  2      &  \\
 Porosity $\phi$          &  0.4      &  \\
 Ice/dust ratio  $\chi_0$ &  0.1$\sim$0.2      &  \\
% Heat conductivity      & $\sim5\times10^{-3}$  & \\
 \hline
\end{tabular}
\end{table}

\begin{figure}
\includegraphics[scale=0.58]{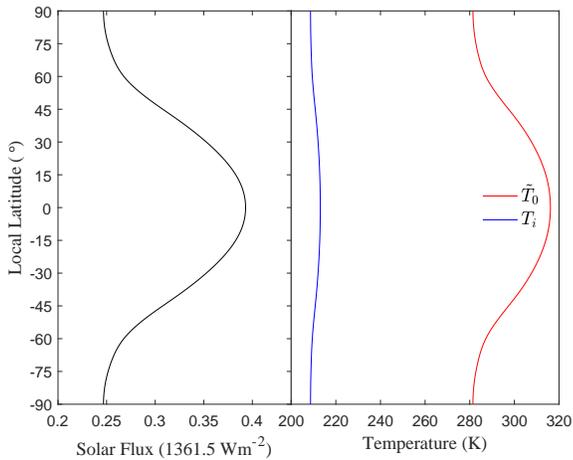}
  \centering
  \caption{Left panel: Annual mean solar insolation as a function of local latitude on Phaethon;
  Right panel: Equilibrium subsurface temperature $\tilde{T}_0$ and ice front temperature
  $T_{\rm i}$ of Phaethon, assuming a geometric Albedo of $p_{\rm v}=0.122$.
  }\label{PhAFAT}
\end{figure}

The equilibrium subsurface temperature $\tilde{T}_0$ and the ice front temperature $T_i$
calculated from our model are shown in Figure \ref{PhAFAT}. The ice front temperature $T_i$
($208 - 213\rm~K$) is found significantly lower than the equilibrium subsurface temperature
$\tilde{T}_0$ ($281 - 316\rm~K$), indicating a larger temperature gradient in the dust mantle
than we found for Pallas and thus a larger rate of energy consumption by ice sublimation.

As Figure \ref{PhAFAT} shows, the ice front temperature $T_i$ varies little with latitude.
Considering that the rotation axis may have changed randomly over Phaethon's dynamical
history \citep{Hanus2016}, it is thus likely that the latitudinal variation of the ice
front temperature has not had any particular significance. Therefore, we use an average
ice front temperature of
\[T_{\rm i}\sim210\rm~K.\]
in the following.
The inward retreating rate of the ice front can then be estimated to be
\[R_r\sim5.68\times10^{-9}\rm~m^2s^{-1},\]
along with the time of losing all ice of
\[t_{\rm L}\sim\frac{R^2}{6R_r}\sim6\rm~Myr.\]

\begin{figure}
\includegraphics[scale=0.58]{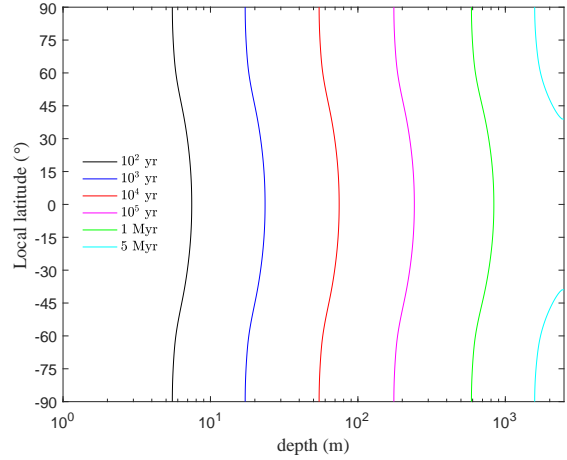}
  \centering
  \caption{Phaethon: Secular inward motion of the ice front.
  }\label{PhHt}
\end{figure}

As a consequence, if Phaethon was to retain a significant fraction of its initial ice,
then it must have been in its present orbit less then about 6 Myr. If the asteroid is a
fragment of Pallas, then as \citet{Todorovi2018} shows, its orbit may have evolved from
a 5:2 or a 8:3 mean motion resonance with Jupiter to its present orbit in $\sim0.294\rm~Myr$.
Figure \ref{PhHt} shows that the depth to the ice front can be $<600$ m if Phaethon arrived
at its present orbit no longer than $1\rm~Myr$ ago. Therefore, it is totally possible that
Phaethon still retains ice buried in its interior.

\subsection{Mean water gas production rate}
Assuming ice in the interior of Phaethon, we can predict the gas production rate,
which would be the key parameter to relate our model to observation. With an ice front
temperature of $T_{\rm i}\sim210\rm~K$, the sublimation rate $J$ at the ice front
can be related to the depth to the ice front $h_{\rm i}$ using Equation (\ref{Jhi}).
The total production rate $J_t$ can then be estimated using
\begin{equation}
J_t=4\pi (R-h_{\rm i})^2J.
\label{Jt}
\end{equation}

\begin{figure}
\includegraphics[scale=0.58]{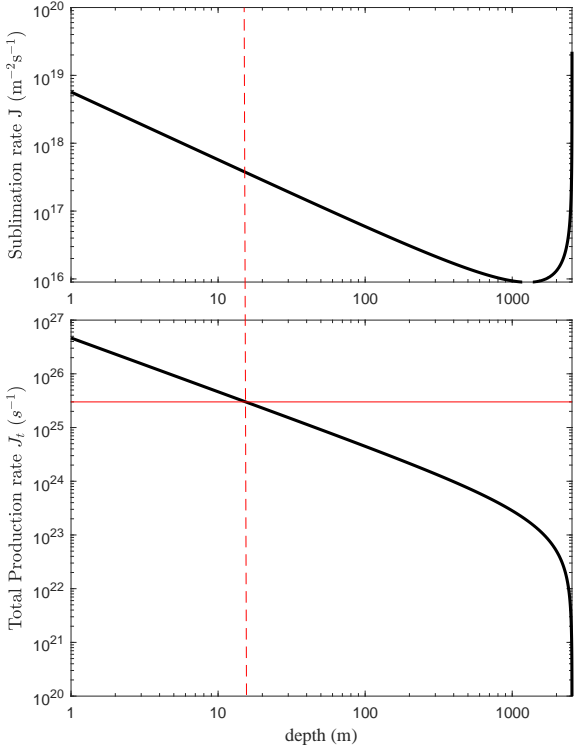}
  \centering
  \caption{Phaethon: Upper panel: Sublimation rate $J$ at the ice front as a function of the
  depth to the ice sublimation front; Lower panel: Total production rate $J_t$.
  }\label{PhnJ}
\end{figure}

In the upper panel of Figure \ref{PhnJ}, we show how the mean sublimation rate $J$
at the ice front relates to the depth to the ice front $h_{\rm i}$. The sublimation
rate $J$ initially decreases with $h_{\rm i}$ increasing from zero to $R/2$. This is
caused by the heat flow (and thereby the sublimation rate) decreasing with increasing
$h_{\rm i}$ at approximately constant temperature difference between the surface and
the sublimation front ($\tilde{T}_0$-$T_i$). At depths greater than $R/2$, the spherical
geometry effect begins to dominate and causes the heat flow per unit area on the sublimation
front to increase as ($R-h_{\rm i}$) decreases. The sublimation rate increases along with
the heat flow and the remainder of the ice near the center will be lost quickly.

The lower panel of Figure \ref{PhnJ} shows the mean total gas production rate $J_t$ as a
function of depth to the sublimation front. Given observational constraints, we could use
the figure to estimate the thickness of the dust mantle above the sublimation front.
However, up to date, no ground-based telescope has ever observed a gas coma around Phaethon.
Thus the total water gas production $J_t$ from Phaethon must be
\[J_t<\sim3\times10^{25}\rm~s^{-1}\]
(Table III in \citet{Chamberlin1996}), implying that the present-day depth to the ice
front should be at least $\sim15\rm~m.$

On the other hand, according to Figure \ref{PhHt}, the ice front can move from $\sim1$ m
depth to $\sim20$ m in about $\sim10^3\rm~yr$. Thus, if we assume the ideal case that the
ice front is at 15 m presently, then it is possible that
\[h_{\rm i}< 1\rm~m\]
for Phaethon $\sim10^3$ yr ago, and Phaethon may come to its present orbit about
$\sim10^3\rm~yr$ ago.

The long-term model will not be applicable for $h_{\rm i}< 1\rm~m$ because differences to
the short-term thermal cycle would be significant. Nevertheless, we can argue that the
total water production rate could have been $J_t>10^{27}\rm~s^{-1}$,
which would be enough to supply the Geminids stream.

\section{Short-term thermal cycle}
\subsection{Seasonal and diurnal temperature cycle}
Considering the large eccentricity of Phaethon's orbit, it is expected that the surface
temperature will vary considerably between perihelion and aphelion.

\begin{figure}
\includegraphics[scale=0.58]{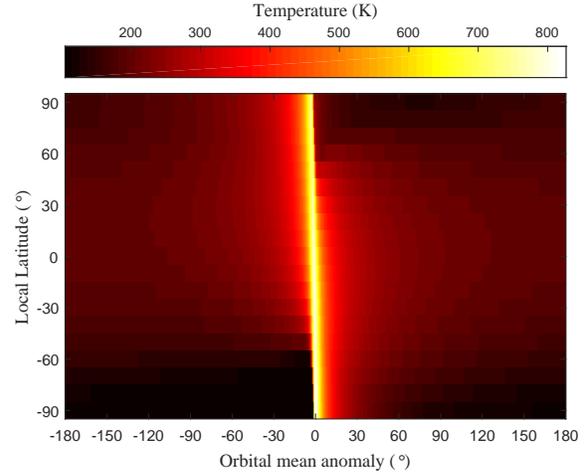}
  \centering
  \caption{Phaethon: Along-orbit variation of the diurnally averaged surface temperature
  as a function of local latitude.
  }\label{Phsst}
\end{figure}

In Figure \ref{Phsst}, we present a map of Phaethon's surface temperature as a function
of local latitude and mean anomaly. The temperature plotted has been averaged over one
rotational period, each. As Figure \ref{Phsst} shows, the surface temperature increases
sharply near perihelion, in particular for the South polar region($-50^\circ$ to $-90^\circ$
latitude), while for the North polar region temperature increases more gradually but decreases
rapidly after perihelion passage.

\begin{figure}
\includegraphics[scale=0.41]{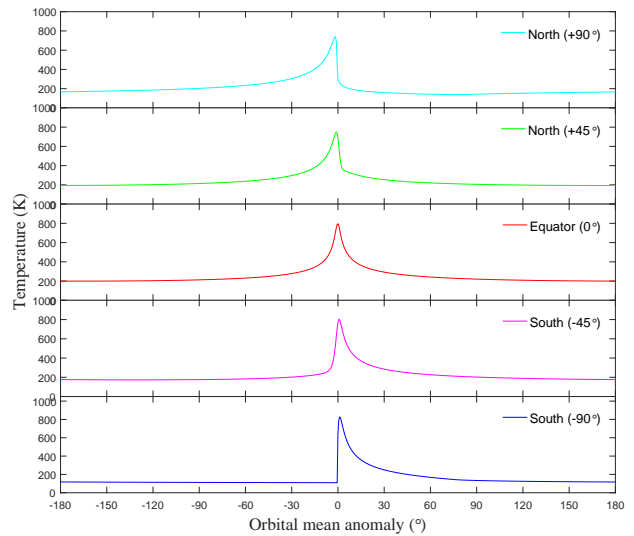}
  \centering
  \caption{Along-orbit variation of the diurnally averaged surface temperature at the North pole, the Equator
  and the South pole of Phaethon.
  }\label{Phstv}
\end{figure}

\begin{figure}
\includegraphics[scale=0.41]{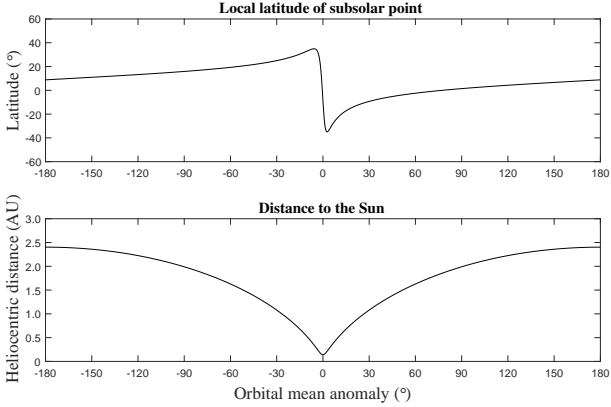}
  \centering
  \caption{Upper panel: Along-orbit variation of the latitude of the sub-solar point on Phaethon;
  Lower panel: Along-orbit variation of the heliocentric distance to Phaethon.
  }\label{Phssds}
\end{figure}

The difference in temperature between the two polar regions is also shown in Figure \ref{Phstv}
and compared with the temperature variation at mid-latitude and at the Equator. The seasonal
effect is significant at all latitudes, causing temperature variations $>600$ K, but is strongest
at the South Pole where temperature rapidly increases by $\sim700$ K in about $2\sim3$ days near
perihelion. These temperature variations can be explained as the results of the swift shift of
the sub-solar point from the northern to the southern hemisphere along with the rapid decrease
of the heliocentric distance near perihelion (compare Figure \ref{Phssds}).

\begin{figure}
\includegraphics[scale=0.41]{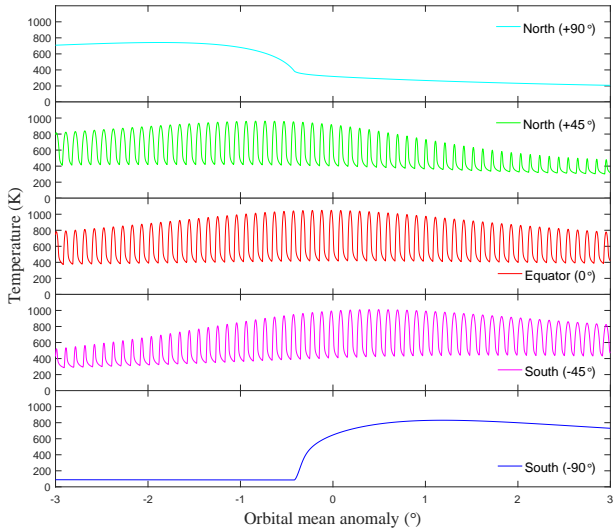}
  \centering
  \caption{Surface temperature variation at the North pole, the Equator and the South pole near
  perihelion for mean anomaly values between $-3^\circ$ and $+3^\circ$ showing both the effects
  of rotation and orbital motion.
  }\label{Phnpsdtv}
\end{figure}

Figure \ref{Phnpsdtv} shows the variation of the surface temperature near perihelion in more detail
including variations caused by rotation. As expected, rotation effects can be neglected at the poles
but become significant at low latitudes where the diurnal temperature variation can also be as large
as $\sim600$ K.

\subsection{Sublimation/condensation cycle near the South pole}
\begin{figure}
\includegraphics[scale=0.51]{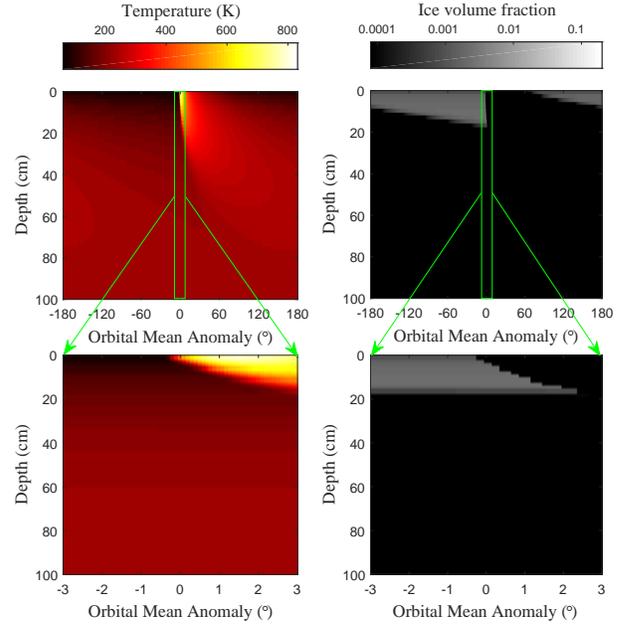}
  \centering
  \caption{Simulation of a Sublimation/condensation cycle on the South pole of Phaethon.
  The initial ice/dust mass ratio $\chi_0$ is assumed to be $0.15$ (corresponding to initial
  ice volume fraction $\sim0.3$) and the initial dust mantle thickness $h_{\rm i}=15$ m.
  }\label{scycle}
\end{figure}

According to Figure \ref{Phstv}, the temperature at the South polar region will be
sufficiently high to drive sublimation only between about $0^\circ$ and $60^\circ$ mean
anomaly, equivalent to $\sim1/6$ of the orbit. For the other $\sim5/6$ of the orbit, the
surface temperature keeps lower than $~100$ K. Any out-flowing gas will condense in a
near-surface layer of $\sim l_{\rm sst}/5\approx20\rm~cm$ thickness, and the condensed
ice will reach a volume fraction of $\sim0.2\%$, there. This is shown in Figure \ref{scycle}
where a simulated cycle of sublimation and condensation is presented.

From Figure \ref{scycle}, we further find that the condensed ice will be quickly
sublimated near perihelion within $\sim1/180$ of the orbital period $(\sim2.9$ days).
This transient sublimation pulse can generate a comparatively large gas outflow,
$J_{\rm s,max}$, which can be estimated using
\[J_{\rm s,max}\sim180\times\frac{5}{6}\tilde{J}_{\rm s}=150\tilde{J}_{\rm s},\]
where $\tilde{J}_{\rm s}$ is the annual mean outflow water flux.

Considering the upper-limit case discussed in section 3.3,
\[\tilde{J}_{\rm s}<\frac{3\times10^{25}\rm~s^{-1}}{4\pi R^2}\sim3\times10^{17}\rm~m^{-2}s^{-1}\]
for present Phaethon and
\[\tilde{J}_{\rm s}\sim\frac{10^{27}\rm~s^{-1}}{4\pi R^2}\sim10^{19}\rm~m^{-2}s^{-1}\]
$10^3\rm~yr$ ago. The maximum outflow water flux near perihelion is then
$J_{\rm s,max}<\sim4.5\times10^{19}\rm~m^{-2}s^{-1}~\rm$ at~present
and
$J_{\rm s,max}\sim1.5\times10^{21}\rm~m^{-2}s^{-1}$ a thousand years ago.

\subsection{Dust Acceleration}
The outflow gas can blow away small dust particles, but the important question is,
how large a particle can be lifted off the surface? To answer this question,
we proceed to compare the gravity force acting on a particle to the gas drag force.

Assume a spherical dust particle with effective radius $b_{\rm d}$ and mean density
$\rho_{\rm d}$. Then, considering rotation, the gravity force on a dust particle at
local latitude $\theta$ can be expressed as
\begin{equation}
F_{\rm G}=\frac{4}{3}\pi b_{\rm d}^3\rho_{\rm d}
\left(R\omega^2\cos^2\theta-\frac{GM}{R^2}\right),
\end{equation}
where $R$ is the radius of the host body and $\omega$ its rotation angular velocity.

The gas drag force on the dust particle can be estimated from:
\begin{equation}
F_{\rm drag}=\frac{{\rm d}P_{\rm M}}{{\rm d}t}\sim
\frac{\hat{m}\tilde{v}_{\rm th}\cdot C_{\rm d}\pi b_{\rm d}^2\cdot J_{\rm s}{\rm d}t}{{\rm d}t}
=C_{\rm d}\pi b_{\rm d}^2\hat{m}\tilde{v}_{\rm th}J_{\rm s},
\end{equation}
where $P_{\rm M}$ is the momentum of a dust particle, $C_{\rm d}$ the drag
coefficient, $\tilde{v}_{\rm th}$ the thermal velocity of the gas molecules at the surface,
and $J_{\rm s}$ is the outflow number flux. Thus the critical size of a dust particle that can
be lifted off by the gas flow can be estimated from
\[0=F_{\rm G}+F_{\rm drag},\]
giving
\begin{equation}
b_{\rm max}\sim\frac{\hat{m}v_{\rm th}J_{\rm s}}{\rho_{\rm d}
\left(\frac{GM}{R^2}-R\omega^2\cos^2\theta\right)}
\end{equation}
if using $C_{\rm d}\sim4/3$ \citep{Epstein1924,Podolak2016} for spherical dust particles.

For present-day Phaethon, considering a maximum surface temperature of $\sim800$ K
near perihelion, but a steady mean gas flux of
\[\tilde{J}_{\rm s}<\sim3\times10^{17}\rm~m^{-2}s^{-1},\]
the critical radius $b_{\rm max}$ is only about $1.8~\mu m$.
However, if considering the transient strong gas flow near perihelion,
\[J_{\rm s,max}<\sim4.5\times10^{19}\rm~m^{-2}s^{-1},\]
we find $b_{\rm max}$ to be about $\sim260\rm~\mu m$, still assuming a maximum
surface temperature of $\sim800$ K.
In addition, for Phaethon $10^3\rm~yr$ ago, we can easily estimate $b_{\rm max}$
to be as large as $1\rm~cm$. These estimates are well consistent with the typical
sizes ($\sim10~\mu\rm{m}$ to $\sim1\rm~cm$) of the Geminid stream particles.

Our model predicts that for present-day Phaethon the transient gas outburst will
only occur at the south polar region and that it will last only for a short time
of ($<P_{\rm orbit}/180\sim2.9$ days). Large dust particles can be blown away
during perihelion passage only, for Mean Anomaly between $0^\circ$ and $2^\circ$.
This is consistent with observation, since the dust tails reported by \citet{Jewitt2013}
and \citet{Hui2017} appeared at Mean Anomaly between $0^\circ$ and $1^\circ$ and
lasted only $\sim2$ days.

\subsection{Thermal stess near perihelion}
Large temperature variation can generate strong thermal stress, probably causing
fracture of rocks or boulders. Using Hooke's law, we can estimate the thermal
stress via
\begin{equation}
\sigma_{\rm th}=E\Delta l=E\alpha_{\rm th}\Delta T~,
\end{equation}
where $E$ is Young's modulus, and $\alpha_{\rm th}$ is the linear coefficient
of thermal expansion.

We estimate the thermal stress that will be generated by a temperature variation
$\Delta T>600$ K for pyroxene and plagioclase as two representative examples.
Their mechanical and thermal properties are well-known and are listed in Table \ref{mapa}.

The thermal stress induced by a temperature variation of $\Delta T>600$ K will
be $\sigma>840$ MPa for pyroxene and $\sigma>204$ MPa for plagioclase. The
fatigue strength of typical rock has been measured to be on the order of
$<200$ MPa \citep{Krokosky1968}. Thus, a temperature variation of $\Delta T>600$ K
on the surface of Phaethon during each perihelion passage can be expected to cause
cracks in typical rock. These cracks can propagate into the rock due to subsequent
continuous thermal cycling and may lead to fracture and the production of dust
particles over the course of time. However, this mechanism of dust formation will
occur on time scales longer than the seasonal cycle of sublimation and condensation.
Therefore, the dust particles on the surface that can be blown away to form an
observable tail are expected to become gradually exhausted with time. The high
average surface thermal inertia $\sim600$ $\rm Jm^{-2}s^{-0.5}K^{-1}$ of Phaethon
\citep{Hanus2016} is similar to that of Itokawa, suggesting that there should
be little dust on the surface of the asteroid. The cycle of the appearance of the
tail would then be predicted to depend more on the time scale of dust formation
than gas release.

\begin{table}\footnotesize
 \renewcommand\arraystretch{2}
 \caption{Material Properties for Pyroxene and Plagioclase \citep{Molaro2015}.}
 \label{mapa}
 \begin{tabular}{@{}lccc@{}}
 \hline
  Properties & Symbol & Pyroxene & Plagioclase\\
 \hline
  Young's modulus(GPa)  &   $E$   & $175$ & $85$ \\
  Possion's ratio       & $\gamma$ & $0.23$  & $0.33$ \\
  linear expansion coefficient $K^{-1}$ & $\alpha$ & $0.8\times10^{-5}$ & $0.4\times10^{-5}$  \\
 \hline
\end{tabular}
\end{table}

\section{Discussion and Conclusion}
Because Phaethon is hitherto the only small body that is dynamically associated with the
Geminid stream, it has been widely accepted as its parent body although the mechanism of
how Phaethon would supply the Geminid stream has remained unclear. Here, on the basis of
theoretical considerations, we find that Phaethon could have retained ice buried underneath
the surface. We found that the tails reported in \citet{Jewitt2013} and \citet{Hui2017}
can be explained by transient gas outbursts near perihelion due to cycles of sublimation
and condensation. The proposed mechanism would be able to supply the Geminid stream for
the past $\sim1000$ years, even though a comet-like coma has not been detected. The model
implies that Phaethon should have been transferred to its present orbit $\sim1000$ years ago.

The feasibility of the proposed model depends on the validity of two assumptions:
(1) Phaethon comes from the main-belt as an initially icy body, possibly a fragment
of Pallas; (2) Phaethon was delivered from the main-belt to its present orbit about
$\sim1000$ years ago. Although these two assumptions have support from some previous
works, there exist other arguments against the assumptions. For example, \citet{Hanus2016}
did a backward orbit calculation of Phaethon for the past 1 Myr. They find only slight
changes $\sim10\%$ of the semi-major axis, suggesting that Phaethon may have remained
in its present orbit for several Myrs. \citet{Reddy2018} found the spectra of Phaethon
obtained during the night of 12 December 2017 to be featureless in the $3~\mu m$ region,
suggesting that its surface is not hydrated, whereas Pallas exhibits a sharp $3~\mu m$ band.
Still, the data of \citet{Reddy2018} cannot rule out that the surface of Phaethon is
covered by a thick dry dust mantle below which ice may exist.

On the basis of our modeling, we conclude:

(1) It is possible, for present-day Phaethon, to still retain buried water ice
in its interior; the dry dust mantle should have a thickness of $>15$ m at least.
Sublimation of the buried ice is too weak to generate a coma observable with
present ground-base instruments.

(2) A thousand years before present, Phaethon's dust mantle could have had a
thickness $<1$ m, which would have allowed Phaethon to be sufficiently active
to supply the Geminid stream up to the present. If true, Phaethon should have
been transferred from the main belt to its present orbit about $10^3$ years ago.

We propose the following mechanisms to explain the relation to the Geminid stream
and the observed dust tail:

(1) Sublimation of ice below the growing dust mantle could have provided the Geminid
particles during the past millennium. A significant sublimation/condensation cycle is
predicted for Phaethon's south polar region even today. The sublimation/condensation
cycle would lead to transient gas outbursts during perihelion passage (Mean Anomaly between
$0^\circ$ and $2^\circ$) capable of blowing away dust particles and explain the observed tail.

(2) The large temperature variation near perihelion could induce sufficiently large
thermal stress to cause fracture of rocks or boulders and would be an efficient mechanism
to produce dust particles on the surface which would then be blown away by out flowing gas
to form the dust tail. But the time scale of this dust producing process should be longer
than the seasonal cycle of water sublimation and condensation and may dominate the cycle
of appearance of Phaethon's tail.

Of course, the proposed mechanisms need to be further examined by new observation and in-situ
detection, especially precise constraints on the total gas production rate, from which we
should be able to estimate how much ice Phaethon really contains. The proposed JAXA/ISAS
DESTINY$^+$ mission to Phaethon would certainly bring new insight to understand the origin
and evolution of Phaethon as well as its connection to the Geminid stream.

\section*{Acknowledgments}
This work was financially supported by the Science and Technology Development
Fund of Macao (Grants No. 119/2017/A3, 061/2017/A2).

\label{lastpage}


\begin{thebibliography}{}

\bibitem[\protect\citeauthoryear{Blaauw}{2017}]{Blaauw2017}
Blaauw, R.C., 2017. The Mass Index and Mass of the Geminid Meteoroid Stream as determined with Radar, Optical and Lunar Impact Data, Planetary and Space Science, 143, 83-88

\bibitem[\protect\citeauthoryear{Boice \& Benkhoff}{2015}]{Boice2015}
Boice, D. C., Benkhoff, J., 2015, Modeling the Near-Sun Object, 3200 Phaethon, 46th Lunar and Planetary Science Conference

\bibitem[\protect\citeauthoryear{Boice}{2017}]{Boice2017}
Boice. D. C., 2017, SUISEI-A Versatile Global Model of Comets with Applications to Small Solar System Bodies, Journal of Applied Mathematics and Physics, 5, 311-320

\bibitem[\protect\citeauthoryear{Borovi\v{c}ka et al.}{2010}]{Borovicka2010}
Borovi\v{c}ka, J., Koten, P., Spurny, P., et al., 2010, Material properties of transition objects 3200 Phaethon and 2003 EH1, Proc. IAU Symposium, 263, 218

\bibitem[\protect\citeauthoryear{Bowell et al.}{1989}]{Bowell}
Bowell, E., Hapke, B., Domingue, D., et al., 1989,  Application of photometric models to asteroids.
In \emph{Asteroids II}, pp. 524-556

\bibitem[\protect\citeauthoryear{Carry et al.}{2010}]{Carry2010}
Carry, B., Dumas, C., Kaasalainen, M., et al, 2010. Physical properties of (2) Pallas, Icarus, 205, 460-472

\bibitem[\protect\citeauthoryear{Chamberlin et al.}{1996}]{Chamberlin1996}
Chamberlin, A.B., McFadden, L.-A., Schulz, R., Schleicher, D.G., \& Bus, S.J., 1996, Icarus, 119, 173

\bibitem[\protect\citeauthoryear{de Le\'{o}n et al.}{2010}]{deLeon2010}
de Le\'{o}n, J., et al., 2010. Origin of the near-Earth asteroid Phaethon and the Geminids meteor shower, A\&A, 513, A26

\bibitem[\protect\citeauthoryear{Epstein}{1924}]{Epstein1924}
Epstein, P.S., 1924. On the resistance experienced by spheres in their motion through gases, Phys. Rev., 23, 710.

\bibitem[\protect\citeauthoryear{Green}{1983}]{Green1983}
Green, S.F., 1983. IAU Circular. No. 3878

\bibitem[\protect\citeauthoryear{Gundlach \& Blum}{2013}]{Gundlach2013}
Gundlach, B., Blum, J., 2013, Icarus, 223, 479-492

\bibitem[\protect\citeauthoryear{Gustafson}{1989}]{Gustafson1989}
Gustafson, B. A. S. 1989, A\&A, 225, 533

\bibitem[\protect\citeauthoryear{Hanu\v{s} et al.}{2016}]{Hanus2016}
Hanu\v{s}, J., et al., 2016, A\&A, 592, A34

\bibitem[\protect\citeauthoryear{Hsieh \& Jewitt}{2006}]{Hsieh2006}
Hsieh Henry H., \& Jewitt D., 2006. A Population of Comets in the Main Asteroid Belt, Science, 312, 561-563

\bibitem[\protect\citeauthoryear{Hui \& Li}{2017}]{Hui2017}
Hui, Man To, \& Li, Jing, 2017. Resurrection of (3200) Phaethon in 2016, Apj, 153,23

\bibitem[\protect\citeauthoryear{Ito et al.}{2018}]{Ito2018}
Ito, T., Ishiguro, M., Arai, T., et al., 2018, Extremely strong polarization of an active asteroid (3200) Phaethon, Nature Communications, 9, 2486

\bibitem[\protect\citeauthoryear{Jenniskens}{2008}]{Jenniskens2008}
Jenniskens, P. 2008, EM\&P, 102, 505

\bibitem[\protect\citeauthoryear{Jewitt}{2013}]{Jewitt2013b}
Jewitt, D. 2013, AJ, 145, 133

\bibitem[\protect\citeauthoryear{Jewitt et al.}{2013}]{Jewitt2013}
Jewitt, D., Li, J., \&, Agarwal, J., 2013, ApjL, 771, L35

\bibitem[\protect\citeauthoryear{Jewitt et al.}{2015}]{Jewitt2015}
Jewitt, D., Hsieh, H., \& Agarwal, J., 2015. In \emph{Asteroids IV} (eds. by P. Michel, F. E. DeMeo,
\& W. F. Bottke), Univ. Arizona Press, Tucson, pp. 221

\bibitem[\protect\citeauthoryear{Krokosky \& Husak}{1968}]{Krokosky1968}
Krokosky, E. M., and Husak, A., 1968. Rock failure under the confined Brazilian test, J.Geophys.Res., 71, 2237-2247

\bibitem[\protect\citeauthoryear{Li \& Jewitt}{2013}]{Li2013}
Li, J., \& Jewitt, D. 2013, AJ, 145, 154

\bibitem[\protect\citeauthoryear{Licandro et al.}{2007}]{Licandro2007}
Licandro J., et al., 2007. The nature of comet-asteroid transition object (3200) Phaethon, A\&A, 461, 751-757

\bibitem[\protect\citeauthoryear{Molaro et al.}{2015}]{Molaro2015}
Molaro, J.L., Byrne, S., \&, Langer, S.A., 2015, JGR, 120, 255-277

\bibitem[\protect\citeauthoryear{Podolak et al.}{2016}]{Podolak2016}
Podolak, M., Flandes, A., Corte, V. D., et al., 2016. A simple model for understanding the DIM dust measurement at comet 67P/Churyumov-Gerasimenko, Planetary and Space Science, 133, 85-89

\bibitem[\protect\citeauthoryear{Reddy et al.}{2018}]{Reddy2018}
Reddy, T.V., Hanu\v{s}, J., Arai, T., et al., 2018. 3-$\mu$m Spectroscopy of Asteroid (3200) Phaethon: Implications for B-Asteroids, 49th LPSC, Contrib. No. 2083

\bibitem[\protect\citeauthoryear{Rivkin et al.}{2002}]{Rivkin2002}
Rivkin, A. S., Howell, E. S., Vilas, F., et al. 2002. In \emph{Asteroids III}, (ed. by W. F. Bottke Jr., A. Cellino, P. Paolicchi, \& R. P. Binzel), University of Arizona Press, Tucson, pp. 235

\bibitem[\protect\citeauthoryear{Sasselov \& Lecar}{2000}]{Sasselov2000}
Sasselov, D.D., \&, Lecar, M., 2000. On the Snow Line in Dusty Protoplanetary Disks, Apj, 528, 995

\bibitem[\protect\citeauthoryear{Schorghofer}{2008}]{Schorghofer2008}
Schorghofer, N., 2008, Apj, 682, 697-705

\bibitem[\protect\citeauthoryear{Todorovi\'{c}}{2018}]{Todorovi2018}
Todorovi\'{c}, N., 2018. The Dynamical Connection Between Phaethon and Pallas, MNRAS, 475, 601-604

\bibitem[\protect\citeauthoryear{Yanagisawa et al.}{2008}]{Yanagisawa2008}
Yanagisawa, M., Ikegami, H., Ishida, M., et al., 2008, Met. Planet. Sci. Suppl., 43, 5169

\end{thebibliography}
\end{document}